\documentclass[reprint,twocolumn,showpacs,preprintnumbers,superscriptaddress,amsmath,amssymb,floatfix, nofootinbib,pra]{revtex4-1}

\usepackage{graphicx}% Include figure files
\usepackage{bm}% bold math
\usepackage{bbold,color}
\usepackage{CJK}
\usepackage{feynmp}
\usepackage{comment}

\newcommand{\mri}{\mathrm{i}}

\newcommand{\Exp}[1]{\mathrm{e}^{\mbox{\footnotesize$#1$}}}
\renewcommand{\vec}[1]{\bm{#1}}
\newcommand{\bs}[1]{{\boldsymbol#1}}
\newcommand{\header}[1]{{\it #1}.-- }
\begin{document}

\begin{CJK*}{UTF8}{gbsn}
\title{Momentum-space dynamics of Dirac quasi-particles in correlated random potentials: interplay between dynamical and Berry phases}

\author{Kean Loon \surname{Lee} (李健伦) }
\affiliation{Centre for Quantum Technologies, National University of Singapore, 3 Science Drive 2, Singapore 117543, Singapore}

\author{Beno\^it \surname{Gr\'emaud}}
\affiliation{Laboratoire Kastler-Brossel, UPMC-Paris 6, ENS, CNRS; 4 Place Jussieu, F-75005 Paris, France}
\affiliation{Centre for Quantum Technologies, %
National University of Singapore, 3 Science Drive 2, Singapore 117543, %
Singapore}
\affiliation{Department of Physics, National University of Singapore, %
2 Science Drive 3, Singapore 117542, Singapore}

\author{Christian \surname{Miniatura}}
\affiliation{INLN, Universit\'e de Nice-Sophia Antipolis, CNRS; 1361 route des Lucioles, 06560 Valbonne, France}
\affiliation{Centre for Quantum Technologies, %
National University of Singapore, 3 Science Drive 2, Singapore 117543, %
Singapore}
\affiliation{Department of Physics, National University of Singapore, %
2 Science Drive 3, Singapore 117542, Singapore}
\affiliation{Institute of Advanced Studies, Nanyang Technological University, %
60 Nanyang View, Singapore 639673, %
Singapore}


\begin{abstract}
{
We consider Dirac quasi-particles, as realized with cold atoms loaded in a honeycomb lattice or in a $\pi$-flux square lattice, in the presence of a weak correlated disorder such that the disorder fluctuations do not couple the two Dirac points of the lattices. We numerically and theoretically investigate the time evolution of the momentum distribution of such quasi-particles when they are initially prepared in a quasi-monochromatic wave packet with a given mean momentum. The parallel transport of the pseudo-spin degree of freedom along scattering paths in momentum space generates a geometrical phase which alters the interference associated with reciprocal scattering paths. In the massless case, a well-known dip in the momentum distribution develops at backscattering (respective to the Dirac point considered) around the transport mean free time. This dip later vanishes in the honeycomb case because of trigonal warping. In the massive case, the dynamical phase of the scattering paths becomes crucial. Its interplay with the geometrical phase induces an additional transient broken reflection symmetry in the momentum distribution. The direction of this asymmetry is a property of the Dirac point considered, independent of the energy of the wave packet. These Berry phase effects could be observed in current cold atom lattice experiments.
}
\end{abstract}

\begin{widetext}
\maketitle
\end{widetext}
\end{CJK*}

\begin{figure}[!htbp]
  \includegraphics[width=0.49\textwidth]{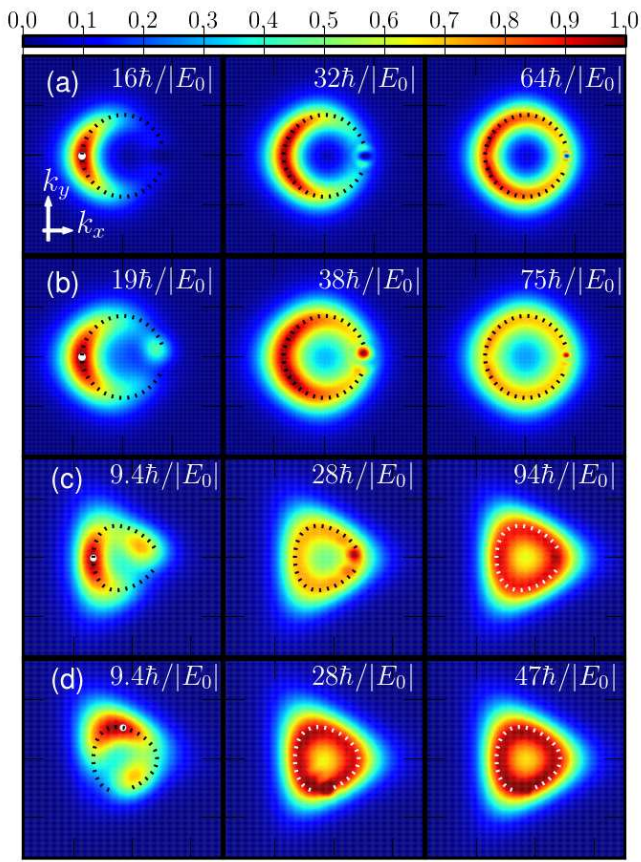}\caption{\label{fig:combine_density}(Color online) Snapshots of the diffuse momentum-space density $n_D(\vec{q}, t)$ (normalized to its maximum value) obtained at different times when Dirac quasi-particles are subjected to a correlated disorder with strength $W$ and correlation length $\zeta$. The initial momentum (white disks) is $\vec{k}_0 = \vec{K}+\vec{q}_0$ with $\vec{q}_0=-q_0\hat{\vec{e}}_x$, $q_0 >0$ and the initial energy is chosen in the lower band, $E_0=-\varepsilon_{\vec{q}_0}$. The black and white dotted lines mark the energy shell, but leave out the region around $-\vec{q}_0$ for clarity. (a) $\pi$-flux square lattice with $m=0$, $q_0\zeta = 2$, $q_0a=0.41$, $W=0.29|E_0|$, $|E_0|\tau_s/\hbar=3.1$ and $|E_0|\tau_B/\hbar=25$. The energy shell is filled symmetrically with respect to the $x$-axis and develops a CBS dip at $-\vec{q}_0$ due to a $\pi$ Berry phase. (b)  $\pi$-flux square lattice with $mc^2=0.53|E_0|$, $q_0\zeta = 2$, $q_0a=0.41$, $W=0.24|E_0|$, $|E_0|\tau_s/\hbar=2.4$ and $|E_0|\tau_B/\hbar=21$. The energy shell is filled anisotropically at short times due to the interplay between dynamical and geometrical phase factors.
  %CBS peak develops at $-\vec{q}_0$ for large enough mass. 
(c) Graphene lattice with $mc^2=0.53|E_0|$, $q_0\zeta \approx 1.27$, $q_0a=0.25$, $W=0.37|E_0|$, $|E_0|\tau_s/\hbar=2.35$ and $|E_0|\tau_B/\hbar=5.9$. Due to trigonal warping, the energy shell is no longer inversion-symmetric: interfering reciprocal paths are always off-shell and the observed fringe structure disappear in time. (d) Same as (c) except that $\vec{q}_0=q_0\hat{\vec{e}}_y$ where $q_0\zeta=1.34$ and $q_0a=0.27$. It shows that the transient asymmetry observed at a given energy around a given Dirac point does not depend on the polar angle of the initial momentum.}
 \end{figure}

\header{Introduction}Topology and gauge fields, as exemplified by Chern numbers~\cite{chern1946}, Berry phases and curvatures~\cite{berry1984}, are key concepts at the heart of many condensed-matter phenomena~\cite{ttkn1982,xiao2010}. One main reason is that Bloch's theorem introduces wave numbers ${\bf k}$ belonging to a parameter space with the topology of a torus, the Brillouin zone, and a set of periodic wave functions depending parametrically on ${\bf k}$. Together they offer natural settings for bundles and connections \cite{resta2000}. With the recent advances in the generation of synthetic magnetic fields~\cite{lin2009, cooper2011a,aidelsburger2013} and the ability to load independent or interacting bosons and/or fermions into two-dimensional optical lattices~\cite{aidelsburger2011,jo2012,tarruell2012,*uehlinger2013} where the relevant experimental parameters are perfectly under control, ultracold atoms have become versatile tools to explore directions borrowed from other fields like topological defects \cite{gigla2013}, color superfluidity \cite{rapp2007}, engineering and measuring non-zero momentum-space Berry curvatures~\cite{price2012} or creating quantum Hall states with strong effective magnetic fields~\cite{jaksch2003,mueller2004,sorensen2005,cooper2011b}.

In this work, we consider the momentum space dynamics of Dirac quasi-particles~\cite{fang2003,sundaram1999,fuchs2010}, as found for example in graphene sheets, in the presence of a correlated and weak on-site disorder. A semi-classical approach shows that the dynamics of these quasi-particles is unaffected by the topology of the band structure as long as a net effective electric field is absent~\cite{sinitsyn2005,xiao2010}. We go beyond this semi-classical approach to address the competition happening at short times between Berry phase effects induced by the Bloch states topology and dynamical coherent corrections to transport due to the interference associated with reciprocal scattering paths~\cite{bergmann1984,wolfle2010,akkermans2007}. Usually, these loop interference are constructive and enhance the return probability of a propagating wave, giving rise to weak localization corrections \cite{bergmann1984}. However, it is long known that they are destructive for magnetic or spin-orbit scattering since a rotation of $2\pi$ of the electron spins gives rise to a sign flip of the interference terms. In this case, one rather gets weak {\it anti}-localization effects as revealed by magneto-resistance measurements and suppression of the coherent backscattering (CBS) effect \cite{bergmann1984, aegerter2009}. The very same situation happens in graphene~\cite{suzuura2002,mccann2006,wu2007,neto2009} where the band structure induces two Berry monopoles of opposite charge located at the Dirac points, hence an effective magnetic field in momentum space. The theoretical calculation of the loop interference terms is generally done in the diffusion approximation~\cite{akkermans2007}, i.e. at long times. As a consequence, its dominant contribution is expected to happen in the vicinity of the CBS momentum even in the presence of a Berry phase. We investigate here the short-time regime and show that the momentum-space dynamics exhibits new features, in particular when the Dirac quasi-particles are massive. In this case the dynamics in momentum space is no longer reflection symmetric and an observable fringe pattern develops at a momentum away from the CBS one and moves towards it as time increases. This reflection symmetry breaking is  similar to the shift of interference fringes in a certain direction observed for Aharonov-Bohm experiments in real space. A key aspect of this symmetry breaking is the subtle interplay between dynamical and Berry phases which impacts the \emph{transient} dynamics where off-shell effects contribute significantly, a situation markedly different from the usual steady-state regime. The impact of the dynamical phase is clearly missed in the diffusion approximation. However, momentum space measurements are challenging in condensed-matter systems. From this point of view, cold atom systems offer an interesting alternative as witnessed by the recent observation of CBS with matter waves~\cite{cherroret2012,labeyriel2012,jendrzejewski2012_cbs}. Furthermore, graphene-like experiments can be easily performed with non-interacting cold atoms loaded on a honeycomb~\cite{wunsch2008,lee2009,tarruell2012,*uehlinger2013,lim2013,greif2013}, or on a $\pi$-flux square~\cite{goldman2009,lin2009,aidelsburger2013} optical lattice at an energy close to the charge neutrality point $E=0$ (Dirac regime).

\header{Model}In the following, we consider a cold-atom experiment setup that combines those of Refs.~\cite{labeyriel2012,jendrzejewski2012_cbs,tarruell2012,*uehlinger2013,aidelsburger2013} and focus on the $\pi$-flux square lattice and the honeycomb one, see Appendix \ref{sec:Clean_lattice}. Each lattice has two inequivalent Dirac points located at $\vec{K}=0$ and $\vec{K}' = \frac{\pi}{a} \hat{\vec{e}}_x$ ($\pi$-flux square lattice) and at $\vec{K} = 4\pi/(3\sqrt{3}a) \hat{\vec{e}}_x$ and $\vec{K}' = -\vec{K}$ (honeycomb lattice), where $a$ is the lattice constant. Their common energy is chosen to be $E=0$ (charge neutrality point). For each of these lattices, we start with an initial quasi-monochromatic Gaussian wave packet $|\psi_0\rangle$ with a mean momentum $\vec{k}_0 = \vec{K}+\vec{q}_0$ close to the Dirac point at $\vec{K}$ ($q_0a\ll 1$) and with a momentum spread $\sigma_0\ll|\vec{k}_0|$. This initial state is essentially a plane wave state $|\vec{k}_0\rangle$ and we have checked that decoherence effects due to the finite width of the wave packet~\cite{cherroret2012} are negligible over the times scales considered here. In the following we conveniently choose $\vec{q}_0$ to lie on the negative $x$-axis and the initial energy $E_0$ of $|\vec{k}_0\rangle$ to be negative (lower band).

At time $t=0$, $|\psi_0\rangle$ is subjected to a spatially-correlated disordered potential $V(\vec{r})$ with Gaussian statistics, vanishing mean value $\overline{V}=0$, and Gaussian 2-point correlator $F(\vec{r}_i-\vec{r}_j)=\overline{V(\vec{r}_i)V(\vec{r}_j)}=W^2\exp(-\frac{(\vec{r}_i-\vec{r}_j)^2}{2\zeta^2})$. Here the overbar denotes the ensemble average, $W$ is the disorder fluctuations strength and $\zeta$ is the disorder correlation length. We then numerically compute the wave function $|\Psi(t)\rangle$ at time $t$ in the case where $k_0\zeta \geq 1$. Hereafter, we focus on the diffuse component $\rho_D(t) =\overline{|\delta\Psi(t)\rangle\langle\delta\Psi(t)|}$ where $|\delta\Psi(t)\rangle = |\Psi(t)\rangle-\overline{|\Psi(t)\rangle}$ which is the dominant contribution at times larger than the scattering mean free time $\tau_s(E_0)\equiv \tau_s$ \cite{ballistic}. We have typically used $10^4$ disorder configurations to extract $\rho_D(t)$. The potential fluctuations can only efficiently couple momenta $\vec{k}$ which are typically within $\zeta^{-1}$ away from each other. In particular, when $\zeta \gg a$, inter-valley scattering is suppressed and the neighborhoods of the two inequivalent Dirac points of the clean lattices are then uncoupled \cite{neto2009}. Writing the scattered momenta as $\vec{k} = \vec{K}+\vec{q}$, we thus have $qa\ll 1$. At lowest order in $\vec{q}$ (see Appendix \ref{sec:Clean_lattice}), the dynamics is then governed by the effective Dirac spinor Hamiltonian $H=H_0+V(\vec{r})$, where $H_0= -c (p_x\sigma_x+p_y\sigma_y)+mc^2\sigma_z$ featuring the momentum operator $\vec{p} = -\mathrm{i}\hbar \bs{\nabla}$, the Dirac quasi-particles mass $m$ \cite{mass}, their velocity $c$, and the usual Pauli matrices $\sigma_x,\sigma_y,\sigma_z$. In momentum space and matrix form, $H_0$ reads:
\begin{eqnarray}
  H_0&=&\hbar c \, (-q_x\sigma_x -q_y\sigma_y+q_m\sigma_z) \equiv \hbar c \, \vec{Q}\cdot\bs{\sigma} \nonumber\\
         &=&\underbrace{\sqrt{\hbar^2q^2c^2+m^2c^4}}_{ \displaystyle{ \varepsilon_{\vec{q} }= \hbar c \,Q^2} } \begin{pmatrix}
                                                                                                             \cos\theta_{\vec{q}} & -\sin\theta_{\vec{q}}\Exp{-\mri\varphi_{\vec{q} }}\\
                                                                                                             -\sin\theta_{\vec{q}}\Exp{\mri\varphi_{\vec{q} }} & -\cos\theta_{\vec{q}}
                                                                                                          \end{pmatrix}\!,
\ \label{eq:cleanH}
\end{eqnarray}
where $q_m= mc/\hbar$ is the Dirac particles Compton wave number, $\varepsilon_{\vec{q}}$ their energy dispersion relation and $\bs{\sigma}$ the Pauli vector operator. The spherical angles of $\vec{Q} = (-q_x, -q_y, q_m)$ are $\theta_{\vec{q}}$ and $\varphi_{\vec{q} }+\pi$, $\varphi_{\vec{q} }$ being the polar angle of $\vec{q}$. The spinorial nature of the Dirac Hamiltonian directly reflects the bipartite nature of the graphene lattice. The corresponding spin-up and spin-down basis vectors $|\!\!\uparrow\rangle$ and $|\!\downarrow\rangle$ represents Bloch waves on either one of the two graphene sublattices ($\theta_{\vec{q}} = 0, \pi$). As one may note, the mass term $q_m$, acting as a Zeeman field, lifts the energy degeneracy between these two spin states. More crucially, it also modifies the Berry curvature and leads to the main results of this work (see later discussion).

\header{Numerical parameters}We give here the typical numerical parameters that we have used in our simulations (as shown in Fig.~\ref{fig:combine_density}) and we compare them to their experimental counterparts as found in Refs.~\cite{labeyriel2012,jendrzejewski2012_cbs,tarruell2012,*uehlinger2013,aidelsburger2013}. In our lattice simulation, we have used a tunneling amplitude $J/h\approx600$Hz as typically found in Refs.~\cite{uehlinger2013,aidelsburger2013}. The mass gaps used in our simulations are typically in the range $2mc^2/h\sim 180-300$Hz, while the tunable mass gap achieved in Ref.~\cite{uehlinger2013} is rather around 20Hz. The interesting dynamics in our problem typically occurs at times $t \sim 20-60 \, \hbar/J$. This would mean experimental observation times in the range 200 to 600$\mu$s. In our simulations, the disorder correlation length is $\zeta=5a$. In~\cite{labeyriel2012,jendrzejewski2012_cbs}, one finds $\zeta \sim 200-540$nm. Since $a$ is determined by the laser wavelength used to create the lattice, and is typically in the range $700-800$nm~\cite{uehlinger2013,aidelsburger2013}, this would mean $\zeta \sim 0.3 - 0.7a$. However producing speckle patterns with larger correlation lengths is not difficult. Finally, the disorder strength used in our simulations is $W/h \approx 120$Hz while those reported in Ref.~\cite{labeyriel2012,jendrzejewski2012_cbs} are in the range 1-3kHz. Here again producing a weaker disorder strength is not an experimental issue.

\begin{figure}
\includegraphics[width=0.45\textwidth]{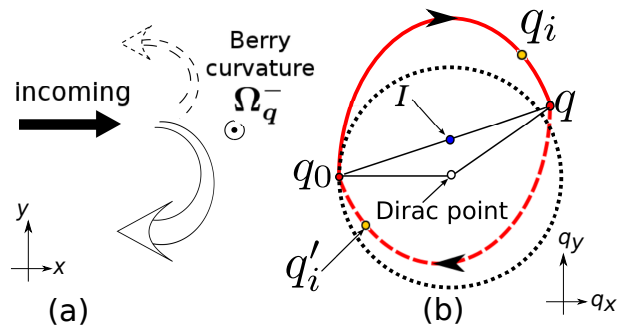}\caption{\label{fig:cross_path_arbitrary} (Color online) (a) Pictorial interpretation of Fig.~\ref{fig:combine_density}b and Fig.~\ref{fig:combine_density}c in terms of real-space scatterings. There is a higher chance for the incoming massive Dirac quasi-particles to be scattered through the bottom (solid larger arrows) half-cicle path than the upper (dashed smaller arrows) half-circle path. (b) Reciprocal scattering paths connecting on-shell momentum $\vec{q}_0$ to final momentum $\vec{q}$ and contributing to the maximally-crossed series $n_C$. The dotted circle represents the energy shell, assumed here to be inversion-symmetric with respect to the Dirac point for simplicity. Given the direct path (solid red), its reciprocal partner (dashed red) is obtained by inversion about $(\vec{q}+\vec{q}_0)/2$ (point I), implying $\vec{q}_i \to \vec{q}'_i= \vec{q}+\vec{q}_0-\vec{q}_i$. Both paths cannot be simultaneously on-shell except when $\vec{q}+\vec{q}_0=0$.
}
\end{figure}

\header{Main results}For weak disorder ($\tau_s\gg \hbar/|E_0|$), inter-band transitions are strongly suppressed as explained from the scattering vertex properties (see Appendix~\ref{sec:maximally_cross}). The band index $s=\pm$ becomes a good quantum number and the dynamics is essentially restricted to the initial energy band. Choosing $E_0<0$, the spinor wave associated with the dynamics is then $|u^{-}_{\vec{q}}\rangle=\sin\frac{\theta_{\vec{q}}}{2}\Exp{-\mri\varphi} |\!\!\uparrow\rangle + \cos\frac{\theta_{\vec{q}}}{2} |\!\!\downarrow\rangle$ 
with eigenenergy $-\varepsilon_{\vec{q}}$. We have duly numerically checked that, with our parameters, the population in the positive energy band is less than three percent of the total population. Figure~\ref{fig:combine_density} shows our main numerical results for the momentum-space diffuse density $n_D(\vec{q},t) = \langle \vec{k} | \rho_D(t)| \vec{k}\rangle$ ($\vec{k}=\vec{K}+\vec{q}$) of a coherent spinorial wave packet propagating in a disordered potential with {\it long-range} spatial correlation. For massless Dirac particles living on the $\pi$-flux square lattice, the dynamics fills symmetrically the energy contours and one witnesses at backscattering the known destructive interference observed in magneto-resistance measurements \cite{bergmann1984}. For massive Dirac particles, the dynamics exhibits a {\it transient} broken reflection symmetry driven by the competition between dynamical and geometric phases acquired by scattering paths in momentum space. We illustrate the meaning of this transient broken symmetry in terms of real-space scatterings in Fig.~\ref{fig:cross_path_arbitrary}a: the incoming Dirac quasi-particles are backscattered by preferentially following clockwise or anti-clockwise U-turn scattering paths depending on the Dirac point considered. The same numerical observations are made for the honeycomb lattice with the important difference that trigonal warping blurs the interference effects as time increases since the energy contours around a given Dirac point are no longer inversion-symmetric.

\header{Theory}At weak disorder, diagrammatic perturbation theory~\cite{akkermans2007} predicts three main contributions to $n_D(\vec{q},t)$: the single scattering background $n_S(\vec{q,t})$ (which we will not address here), the diffusive background $n_L(\vec{q},t)$ associated with multiple scattering paths sequences $\Gamma_N = \{ \vec{q}_0 \to \vec{q}_1 \to (\cdot\cdot\cdot) \to \vec{q}_{N-1} \to \vec{q} \}$ and the maximally-crossed contribution $n_C(\vec{q},t)$ arising from the interference between $\Gamma_N$ and its reciprocal partner $\tilde{\Gamma}_N = \{\vec{q}_0 \to \vec{q}'_1 \to (\cdot\cdot\cdot) \to \vec{q}'_{N-1} \to \vec{q} \}$ where $\vec{q}'_i=-\vec{q}_i+\vec{q}_0+\vec{q}$~(see Appendix \ref{sec:maximally_cross} for derivation), see Fig.~\ref{fig:cross_path_arbitrary}b. The contribution of $N$th-order scattering paths ($N\geq 1$) to $n_D(\vec{q},t)$ is
\begin{eqnarray}
n^{(N)}_{L}(\vec{q},t)&=&\int [d\mu] \, \Exp{-\frac{t}{\tau_s}} \,|\Psi_{\Gamma_N}(t)|^2,
\label{eq:ladderNpath}\\
n^{(N)}_{C}(\vec{q},t)&=&\int [d\mu] \, \Exp{-\frac{t}{\tau_s}} \, \mathrm{Re}(\Psi^*_{\Gamma_N}(t)\Psi_{\tilde{\Gamma}_N}(t)),\label{eq:crossNpath}
\end{eqnarray}
where $[d\mu] = P(\vec{q}_0-\vec{q}_{1})\prod_{i=1}^{i=N-1}\frac{d\vec{q}_i}{(2\pi)^2}P(\vec{q}_{i}-\vec{q}_{i+1})$ is the integration measure with the convention $\vec{q}_{N}=\vec{q}$ and $P(\vec{q})=2\pi\zeta^2W^2\Exp{-\frac{\zeta^2\vec{q}^2}{2}}$ is the Fourier transform of $F(\vec{r})$. Note that the correlation functions $P$ in $[d\mu]$ typically enforce small momentum changes after each scattering, i.e. $|\vec{q}_{i+1}-\vec{q}_i| \lesssim 1/\zeta$. Defining $\omega_i = \varepsilon_{\vec{q}_i}/\hbar$, the amplitude $\Psi_{\Gamma_N}(t)$ associated with the scattering path $\Gamma_N$ is:
\begin{align}
 \Psi_{\Gamma_N}(t)= &\underbrace{\Big[\int \!\!\cdot\cdot\cdot \!\! \int_0^t\;\big(\prod_{i=0}^{i=N} dt_i \Exp{-\mri\omega_it_i}\big) \delta\big(\sum_{j=0}^{j=N}t_j-t\big) \Big]}_{\mathrm{dynamical \ factor}} \nonumber \\
 &\times \underbrace{\Big[\prod_{i=0}^{i=N-1}\langle u^{-}_{\vec{q}_{i+1}}|u^{-}_{\vec{q}_{i}}\rangle\Big]}_{\mathrm{spinor \ factor}}.
 \label{eq:psiN}
\end{align}

To get further physical insight into the problem, we assume $q_0\zeta\gg1$ though this limit is not strictly satisfied for all the data presented here. In this case the momentum changes satisfy $|\vec{q}_{i+1}-\vec{q}_i|\ll |\vec{q}_0|$ and the spinor factor can be conveniently recast into a geometrical phase factor $\prod_{i=0}^{i=N-1}\langle u^{-}_{\vec{q}_{i+1}}|u^{-}_{\vec{q}_{i}}\rangle\approx\exp(\mri\int d\vec{q}\cdot\vec{A}^{-}_{\vec{q}})$, where the Berry connection $\vec{A}^{-}_{\vec{q}}=\mri\langle u^{-}_{\vec{q}}|\nabla_{\vec{q}} u^{-}_{\vec{q}}\rangle$ is the momentum-space analog of a vector potential. We have numerically checked that the dynamical factor vanishes at long times on a time scale set by the spread of intermediate energies (see Appendix \ref{sec:dynamicFac}). The phase of this dynamical factor is well approximated by $(-E_N t/\hbar)$ where $E_N = \frac{1}{N+1}\sum_{i=0}^{i=N}\varepsilon_{\vec{q}_i}$ is the average energy along the scattering path. The amplitude of the dynamical factor forces the various energies contributing to the scattering path to lie close to the energy shell. It can be easily checked that this result is exact for single scattering or in the long-time limit where all intermediate energies are equal. 

\begin{figure}
\includegraphics[width=0.48\textwidth]{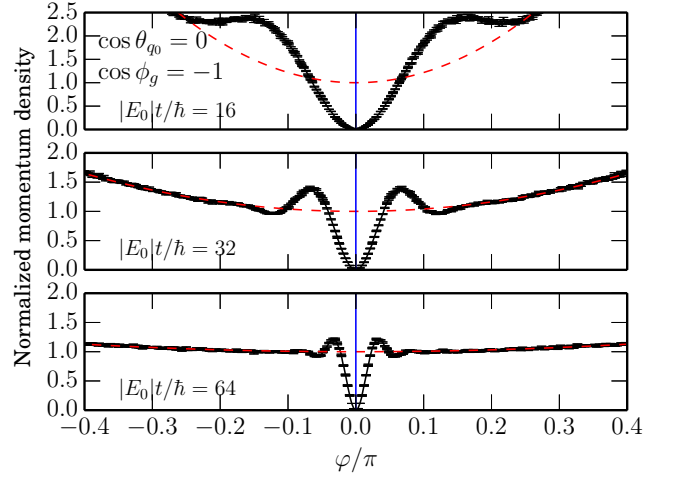}\caption{\label{fig:SO_E-0.8W10g0z5_ring}
(Color online) Normalized momentum-space density $n_D(\vec{q},t)/n_L(q_0,t)$ of massless Dirac particles on a $\pi$-flux square lattice as a function of the polar angle $\varphi$ along the circle $\vec{q}=q_0(\cos\varphi,\sin\varphi)$ at three different times. The simulation parameters are the same as for Fig.~\ref{fig:combine_density}a. To extract the background (red dashes), we use the quadratic fit $\left(1+C(t) \times\varphi^2\right)$ and find $C(t)$ by matching the wings of $n_D(\vec{q},t)$ in the angular range $0.2<|\varphi|/\pi<0.5$. The (blue) vertical line marks the backscattering direction. The dip at backscattering reflects the destructive interference of reciprocal paths amplitudes due to the $\pi$ Berry phase.
}
\end{figure}

As phases factors do not impact $|\Psi_{\Gamma_N}(t)|^2$, $n_L(\vec{q},t)$ fills symmetrically the energy shell with respect to $\vec{q}_0$ in the course of time (see Fig.~\ref{fig:combine_density}). A fully isotropic background $n_{{\rm bg}}(q)$ is obtained after time scales set by the Boltzmann transport time $\tau_B$~\cite{plisson2012}. In the long-time limit, this background is given by:
\begin{align}
n_{{\rm bg}}(q) = \int \frac{dE}{2\pi}\frac{\mathcal{A}(q,E)\mathcal{A}(q_0,E)}{2\pi\nu(E)},
\end{align}
where $\mathcal{A}(q,E) =  2\pi \langle \vec{K}+\vec{q}|\overline{\delta(E-H)}| \vec{K}+\vec{q}\rangle$ is the spectral function and $\nu(E)=\int\frac{d\vec{q}}{(2\pi)^2} \mathcal{A}(q,E)$ is the density of states (DoS). Energy domains where $\nu(E)$ is small contribute more to the background. In our case, it corresponds to states with small $|\vec{q}|$. The maximum of the background is thus slightly shifted from the energy shell $\varepsilon_{\vec{q}} = E_0$ towards the Dirac point, as witnessed in Figs.~\ref{fig:combine_density} and \ref{fig:ring}. On the other hand, we see from Eq.~\eqref{eq:crossNpath}
that $\Gamma_N$ and $\tilde{\Gamma}_N$ combine to form a {\it closed} trajectory $\mathcal{C}$ in momentum space with finite area (Fig.~\ref{fig:cross_path_arbitrary}b). In the large-$N$ limit, this closed trajectory develops a dynamical phase $\phi_d = \Delta t/\hbar$ where
\begin{align}
 \Delta = \lim_{N\to\infty}\!\!\big(E'_N\!\!-\!\!E_N\big) =  \lim_{N\to\infty}\frac{1}{N+1}\sum_{i=0}^{i=N}(\varepsilon_{\vec{q}'_i}-\varepsilon_{\vec{q}_i}),
  \label{eq:dynamicalPhase}
\end{align}
while the geometric phase is
\begin{align}
\phi_g=\oint d\vec{q}\cdot\vec{A}^{-}_{\vec{q}} =\pm \int\!\!\!\!\!\int \! \Omega^{-}_{\vec{q}} \; dS,\label{eq:geometricalPhase}
\end{align}
where the second integral is over a surface $S$ in momentum space with boundary $\mathcal{C}$. The Berry curvature $\bs{\Omega}^{-}_{\vec{q}}=\nabla_{\vec{q}}\times\vec{A}^{-}_{\vec{q}}=\Omega^{-}_{\vec{q}}\vec{e}_z$ is the momentum-space analog of a magnetic field~\cite{fuchs2010}. The sign of $\phi_g$ is positive for anti-clockwise paths and negative for the clockwise paths shown in Fig.~\ref{fig:cross_path_arbitrary}. As a result, $n_L+n_C$ contains modulation terms like $[1+\cos(\phi_d+\phi_g)]$. 

\header{Massless particles: anti-CBS effect}For massless Dirac particles ($m=0$), $\Omega^{-}_{\vec{q}} = \pi \delta(\vec{q})$ and $\phi_g=\pm\pi$ for all paths that enclose the Dirac point. Furthermore, at exact backscattering ($\vec{q}+\vec{q}_0=0$), $\phi_d$ vanishes identically and $n_C$ always gives a destructive contribution resulting in a density dip at $-\vec{q}_0$ (anti-CBS effect), see Figs.~\ref{fig:combine_density}a,~\ref{fig:SO_E-0.8W10g0z5_ring} and~\ref{fig:ring}. Constructive interference peaks can be inferred from the condition $\cos\phi_d=-1$. They are located symmetrically with respect to the $x$-axis and move towards $-\vec{q}_0$ in the course of time, see Figs.~\ref{fig:combine_density}a and~\ref{fig:SO_E-0.8W10g0z5_ring}.

\begin{figure}
\includegraphics[width=0.48\textwidth]{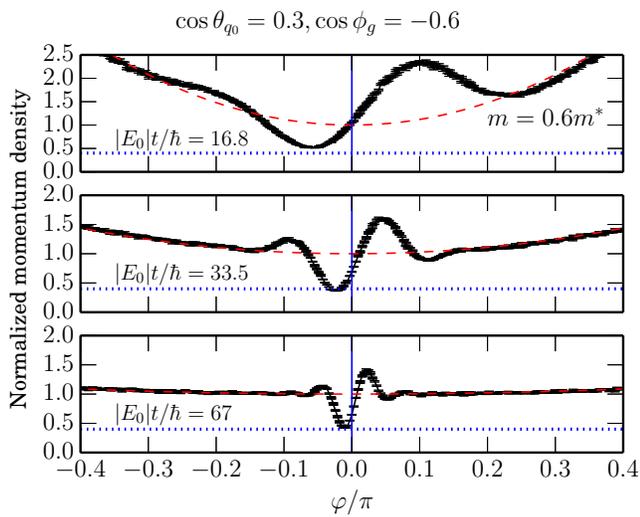}\caption{\label{fig:SO_E-0.8W10g0.5z5_ring}
(Color online) Normalized momentum-space density $n_D(\vec{q},t)/n_L(q_0,t)$ of massive Dirac particles on a $\pi$-flux square lattice as a function of the polar angle $\varphi$ along the circle $\vec{q}=q_0(\cos\varphi,\sin\varphi)$ with $|\vec{q}_0|a=0.41$ at three different times. The rest mass energy is $mc^2=0.3|E_0|$ (equivalently $m = 0.6m^*$ where $m^*c^2=|E_0|/2$, see text) while the other simulation parameters are the same as for Fig.~\ref{fig:combine_density}a. The scattering time is $\tau_s=2.5\, \hbar/|E_0|$. To extract the background (red dashes), we use the quadratic fit $\left(1+C(t) \times\varphi^2\right)$ and find $C(t)$ by matching the wings of $n_D(\vec{q},t)$ in the angular range $0.2<|\varphi|/\pi<0.5$. The blue vertical line marks the backscattering direction. The (blue) dotted horizontal lines give the magnitude of the density ($1+\cos\phi_g$) that is expected in the long time limit at backscattering. As $m<m^*$, $(1+\cos\phi_g) <0$ and the interference between reciprocal paths amplitudes is destructive, giving rise to a dip at backscattering in the long time limit.
}
\end{figure}

\begin{figure}
\includegraphics[width=0.48\textwidth]{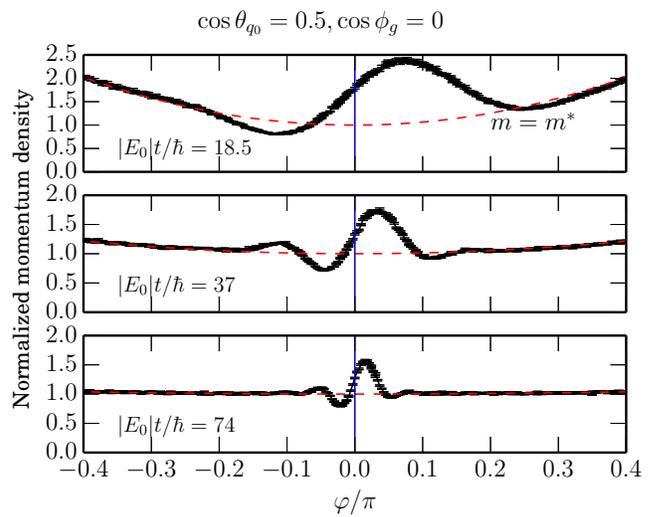}\caption{\label{fig:SO_E-0.8W10g0.924z5_ring}
(Color online) Same as Fig.~\ref{fig:SO_E-0.8W10g0.5z5_ring} but for a rest mass energy $mc^2=0.5|E_0|$ ($m=m^*$, see text). In the long time limit, we expect the interference pattern to disappear since $1+\cos\phi_g=0$.
}
\end{figure}

\begin{figure}
\includegraphics[width=0.48\textwidth]{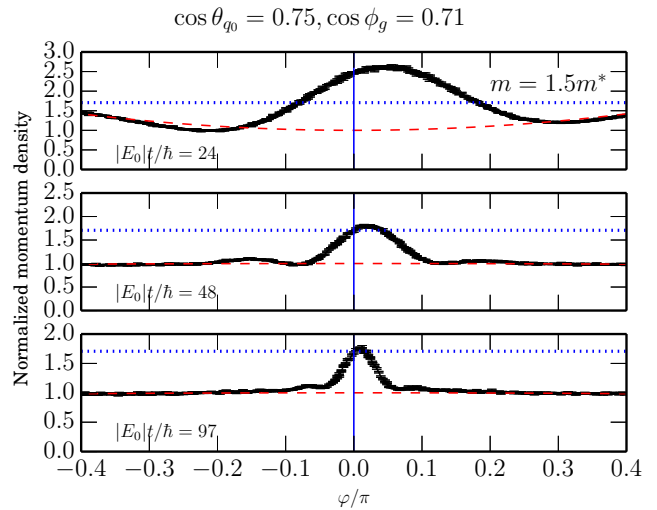}\caption{\label{fig:SO_E-0.8W10g1.814z5_ring}
(Color online) Same as Fig.~\ref{fig:SO_E-0.8W10g0.5z5_ring} but for a rest mass energy $mc^2=0.75|E_0|$ ($m=1.5m^*$, see text). The (blue) dotted horizontal lines give the magnitude of the density ($1+\cos\phi_g$) that is expected in the long time limit at backscattering. As $m>m^*$, $(1+\cos\phi_g) >0$ and the interference between reciprocal paths amplitudes is now constructive, giving rise to a peak at backscattering in the long time limit.
}
\end{figure}

\header{Massive particles: transient broken symmetry}For massive Dirac particles ($m\neq 0$), the picture is different. Now the Berry curvature reads~\cite{fuchs2010}:
\begin{equation}
\Omega^{-}_{\vec{q}} = \frac{\cos\theta_{\vec{q}}}{2Q^2} = \frac{q_m}{2(q^2+q_m^2)^{3/2}}.
\end{equation}
The reflection symmetry is broken and the interference peak on clockwise paths is much greater than the other one, see Figs.~\ref{fig:combine_density}b and~\ref{fig:combine_density}c, as well as Figs.~\ref{fig:SO_E-0.8W10g0.5z5_ring}-\ref{fig:SO_E-0.8W10g1.814z5_ring} for the density as a function of the polar angle on the energy shell. This asymmetry is easily understood since clockwise and anti-clockwise paths have same (positive) dynamical phases but opposite geometrical phases. Thus, as long as $\Omega_0 \neq 2\pi$ (massive case), $\cos(\phi_d-|\phi_g|) > \cos(\phi_d+|\phi_g|)$ which explains the observed asymmetry. One may note that the asymmetry pattern does not depend on the sign of $E_0$ and is thus the same for the two bands above and below the charge neutrality point. For the graphene lattice, the dynamics around the two inequivalent Dirac points display opposite asymmetries because the associated Berry curvature are opposite. We stress that the observed asymmetry cannot be explained if the interfering scattering paths are both on-shell as the dynamic phase of the closed trajectory would identically vanish leaving the interference contribution insensitive to the sign of $\phi_g$.

\begin{figure}
\includegraphics[width=0.48\textwidth]{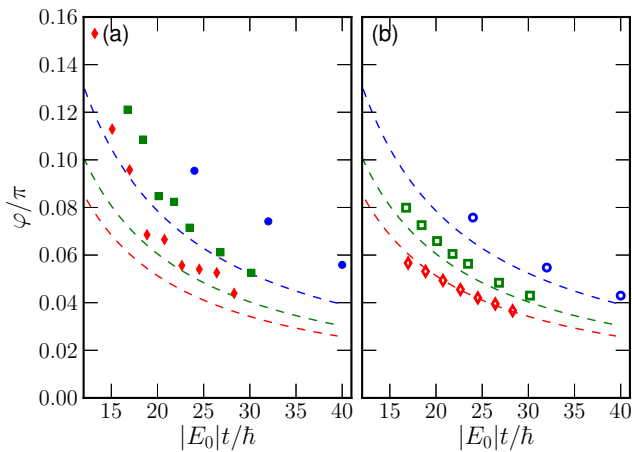}\caption{\label{fig:peakCompare}(Color online) Time evolution of the azimuthal angle $\varphi$ (in units of $\pi$) of the interference peak observed in momentum space for the $\pi$-flux square lattice when $q_0\zeta = 2$, $q_0a=0.41$. Starting from $\varphi = \pi$ (initial momentum $\vec{q}_0$), it decreases and reaches $\varphi = 0$ (CBS momentum $-\vec{q}_0$) in the long-time limit. The full symbols in (a) show the numerical values while the corresponding open symbols in (b) give the theoretical values derived from the ``bright fringe" condition $\cos(\phi_d-|\phi_g|)=1$. Blue circles: $W= 0.29|E_0|$ and $m=0$. In this case, there is another interference peak at $-\varphi$. Green squares: $W=0.27 |E_0|$ and $mc^2=0.3|E_0|$. Red diamonds: $W=0.24|E_0|$ and $mc^2=0.53|E_0|$. Dashed lines are the algebraic predictions computed with Eq.~\eqref{eq:phi_half_circle}. }
\end{figure}

The asymmetry is best seen at times $t$ comparable to $\tau_B$, when the interference peak can be resolved from the diffusive background. Fig.~\ref{fig:peakCompare}a shows the numerically-extracted peak position as a function of time while Fig.~\ref{fig:peakCompare}b shows a rough estimate obtained from the``bright fringe'' condition $\cos(\phi_d-|\phi_g|)=1$, where $\phi_d$ and $\phi_g$ are computed along the path maximizing the product of the disorder correlation function and of the dynamical factor with respect to the intermediate momenta (see Appendix \ref{sec:optimize}). This ``bright fringe" estimate is backed up by an analytical prediction~\cite{dd_half_circle}:
\begin{align}
\label{eq:phi_half_circle}
  \varphi (t) = \frac{\pi^2}{2(1+\cos\theta_{\vec{q}_0})}\frac{\hbar}{|E_0|t}.
\end{align}
The agreement with numerics is qualitatively good, see Fig.~\ref{fig:peakCompare}, with discrepancies by less than a factor $2$ at most.

\begin{figure}[!htbp]
\includegraphics[width=0.48\textwidth]{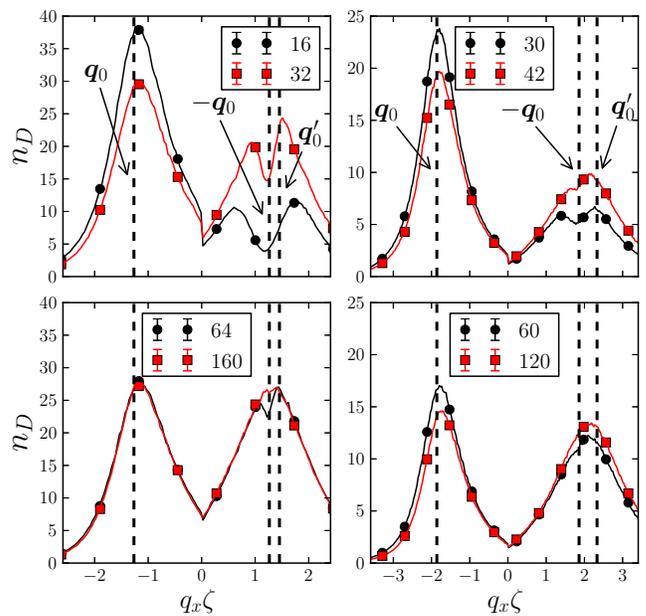}\caption{\label{fig:ring}(Color online)  Diffuse momentum-space density $n_D(\vec{q}, t)$ along the $x$-axis for massless Dirac particles on the honeycomb lattice in the presence of trigonal warping. The times (in units of $\hbar/|E_0|$) are shown next to the symbols in the inserts. Left panel: $q_0\zeta=1.27$, $q_0a=0.25$, $W=0.44|E_0|$, $|E_0|\tau_s/\hbar=2$ and $|E_0|\tau_B/\hbar=9$. The three vertical dashed lines mark, from left to right, the initial momentum $\vec{q}_0$, the CBS momentum $-\vec{q}_0$ and the on-shell momentum $\vec{q}'_0$ opposite to $\vec{q}_0$ with respect to the Dirac point. The CBS dip is centered at $-\vec{q}_0$ but not at $\vec{q}'_0$ as marked by the dashed vertical lines. Right panel: same as left panel but for $q_0\zeta=1.86$, $q_0a=0.37$. One has $|E_0|\tau_s/\hbar=3$ and $|E_0|\tau_B/\hbar=15.6$. In this case the effect of trigonal warping is stronger: $\vec{q}'_0$ is further away from $-\vec{q}_0$ but the CBS dip remains centered at $-\vec{q}_0$. Note that, as it should, the background maximum is located at a momentum slightly lower than $\vec{q}'_0$ but larger than $-\vec{q}_0$ due to DoS effects (see text).}
\end{figure}

\header{Massive particles: interference corrections at backscattering}At backscattering, $\phi_g=\pm \Omega_0/2$, where $\Omega_0 = 2\pi (1-\cos\theta_{\vec{q}_0})$ is the solid angle of the cone subtended by the vector $\vec{Q}_0=(-\vec{q}_0,q_m)$ around the $z$-axis. At a sufficiently long time ($t\gg\tau_B$) the diffusion approximation applies and the diffusive background becomes angularly flat, $n_L = n_{{\rm bg}}(q_0)$. The interference contribution turns constructive when $\Omega_0 \leq \pi$, i.e. for rest mass energies larger than $m_*c^2=|E_0|/2$ (equivalently for Compton wave numbers larger than $m_*c/\hbar = \sqrt{3}q_0$). We thus observe a density \emph{dip} at $\varphi=0$ for $m<m^*$ (see Fig.~\ref{fig:SO_E-0.8W10g0.5z5_ring}) but a \emph{peak} when $m>m^*$ (see Fig.~\ref{fig:SO_E-0.8W10g1.814z5_ring}), both with a magnitude of $1+\cos\phi_g$ relative to $n_L$. At intermediate time ($t\lesssim\tau_B$), we always see a density dip located in the negative-angle sector (respectively a density peak located in the positive-angle sector) moving towards $\varphi=0$ as time increases. This observation is consistent with our analysis of the transient dynamics using the optimized path that maximizes the product of the disorder correlation function and of the dynamical factor (see Appendix \ref{sec:optimize}). However our optimized-path approach incorrectly predicts that the density at backscattering ($\varphi=0$) should have a relative magnitude of $(1+\cos\phi_g)$, which is not what is observed at intermediate times, see e.g. the top two panels in Fig.~\ref{fig:SO_E-0.8W10g0.5z5_ring}. The reason behind the discrepancy is the contribution of other scattering paths. In particular, when $m<m^*$, the optimized path gives a negative contribution to $n_C$ due to a geometrical phase $\phi_g\approx \pm\Omega_0/2$. However the straight line in momentum space connecting $\vec{q}_0$ to $-\vec{q}_0$ is another extremal path and it has zero dynamical and geometric phases. This path gives thus a positive contribution to $n_C$ and impacts the transient dynamics by reducing the magnitude of $n_C$ at backscattering. Surprisingly, these additional scattering paths dress the density dip at negative $\varphi$ with the relative magnitude $(1+\cos\phi_g)$, even at intermediate times (see Fig.~\ref{fig:SO_E-0.8W10g0.5z5_ring}). The moving dip thus features the expected correct dip at backscattering in the long time limit. We have also checked these observations for $m=0.8m^*$ (data not shown).

\header{Effect of trigonal warping on a honeycomb lattice}An astute reader may have noticed the triangular shape of the energy contour in Fig.~\ref{fig:combine_density}c for the graphene case. This trigonal warping~\cite{mccann2006,kechedzhi2007} breaks the inversion symmetry of the energy contours around each Dirac point and prevents $\vec{q}_0$ and $-\vec{q}_0$ to lie on the same energy shell. No direct and reverse paths can then be simultaneously on shell and the CBS dip observed for $m=0$ becomes a transient effect disappearing in the course of time (see Fig.~\ref{fig:ring}). The situation gets worse as the initial energy $E_0$ is chosen further away from the charge neutrality point. Surprisingly enough, this transient CBS dip still occurs at $-\vec{q}_0$ and not at the on-shell point $\vec{q}'_0$ opposite to $\vec{q}_0$. This is confirmed by a path-optimization procedure (see Appendix \ref{sec:CBSpoint}). However a clear understanding of this observation is still missing.

\header{Conclusion}We have highlighted the non-trivial interplay between dynamical and topological phases when Dirac particles are subjected to long-range correlated spatial disorder. This interplay could be experimentally addressed using cold atoms loaded in a $\pi$-flux square or a graphene-like optical lattice and could foster applications in valleytronics~\cite{moldovan2012}.

\begin{acknowledgments}
We gratefully thank Dominique Delande for stimulating discussions. BG and ChM acknowledge support from the LIA FSQL (CNRS). The Centre for Quantum Technologies is a Research Centre of Excellence funded by the Ministry of Education and the National Research Foundation of Singapore.
\end{acknowledgments}

\appendix

\section{\label{sec:Clean_lattice}Clean lattice Hamiltonians}
\begin{figure}
  \includegraphics[width=0.35\textwidth]{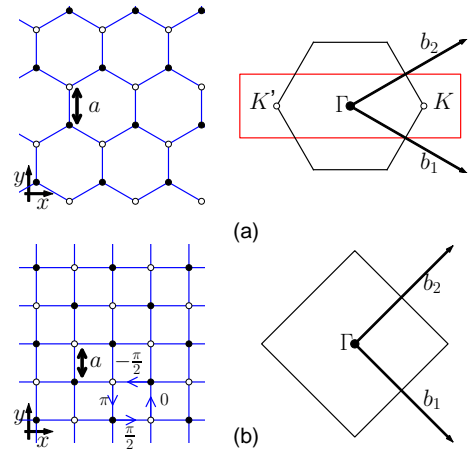}\caption{\label{fig:lattice} (a) Honeycomb lattice and its hexagonal and rectangular (in red) Brillouin zone. The two inequivalent Dirac points are located at $\vec{K}= (\vec{b}_1+\vec{b}_2)/3 = \frac{4\pi}{3\sqrt{3}a} \hat{\vec{e}}_x$ and at $\vec{K}' = -\vec{K}$ where the $\vec{b}_i$s are the reciprocal lattice vectors and $a$ the lattice constant. (b) $\pi$-flux square lattice with its accumulated phase factors around a plaquette and its square Brillouin zone. There is a Dirac point located at the band centre $\Gamma$ at $\vec{K}=0$ and at the band edge at $\vec{K}'=(\vec{b}_1+\vec{b}_2)/2=\frac{\pi}{a} \hat{\vec{e}}_x$. Both lattices are bipartite with $\textsc{a}$-type (full circles) and $\textsc{b}$-type (open circles) sites, each sublattice being a Bravais lattice.
  }
\end{figure}

Our numerical simulations have been carried out using a tight-binding model on either a $\pi$-flux square lattice or on the honeycomb lattice, see Fig.~\ref{fig:lattice} for our conventions. These two lattice models lead to the same low-energy effective Dirac Hamiltonian given by~Eq.~(1) in main text. The tight-binding Hamilton operator is given by:
\begin{equation}
  H_{\rm lat}=-J\sum\limits_{\langle i,j\rangle} \left(\Exp{\mri A_{ij}}c^\dag_i c_j +\mathrm{h.c.}\right) + \delta (\sum_{i\in\textsc{a}}n_i -\sum_{j\in\textsc{b}}n_j),
\end{equation}
where $J\geq 0$ is the hopping amplitude and $2\delta$ is the on-site energy imbalance between nearest-neighbor sites. Here $c^\dag_i \,  (c_i)$ is the creation (annihilation) operator at lattice site $i$, $n_i=c^\dag_i c_i$ is the corresponding number operator, \emph{h.c.} means Hermitian conjugate and $\langle i,j\rangle$ restricts the summation to nearest-neighbor pairs. $A_{ij}$ is identically zero for the honeycomb lattice whereas $\sum_{\Box} A_{ij} = \pi$ for the $\pi$-flux square lattice, where $\Box$ denotes an elementary anti-clockwise plaquette.

Defining the creation operator $f_{\vec{k},\sigma}^\dag$
of Bloch states living on sublattices $\sigma=\textsc{a},\textsc{b}$ through
\begin{subequations}
\begin{align}
  f_{\vec{k}\sigma}^\dag &= \frac{1}{\sqrt{N_c}}\sum_{i\in \sigma}\Exp{\mri \vec{k}\cdot\vec{r}_i} c^\dag_i,\\
   c_{i\in\sigma}^\dag &= \frac{1}{\sqrt{N_c}}\sum_{\vec{k}\in\Omega}\Exp{-\mri \vec{k}\cdot\vec{r}_i} f^\dag_{\vec{k}\sigma},
\end{align}
\end{subequations}
where $N_c$ is the number of unit cells and $\Omega$ the Brillouin zone, the lattice Hamilton operator can be recast as
\begin{equation}
 H_{\rm lat}=
 \sum_{\vec{k}\in \Omega}(f_{\vec{k},\textsc{a}}^\dag, f_{\vec{k},\textsc{b}}^\dag)\begin{pmatrix}
               \delta & Z_{\vec{k}} \\
               Z_{\vec{k}}^* & -\delta
           \end{pmatrix}\begin{pmatrix}f_{\vec{k},\textsc{a}}\\f_{\vec{k},\textsc{b}}\end{pmatrix},
\end{equation}
where
\begin{equation}
 Z_{\vec{k}}= -J\Exp{-\mri\frac{k_ya}{2}}\Bigl(\Exp{\mri\frac{3k_ya}{2}}+2\cos\bigl(\frac{\sqrt{3}k_xa}{2}\bigr)\Bigr)
 \end{equation}
for the honeycomb lattice and
 \begin{equation}
Z_{\vec{k}}=-2J\bigl(\sin(k_xa)-\mri\sin(k_ya)\bigr)
\end{equation}
for the $\pi$-flux square lattice. The two inequivalent Dirac points are found by solving $Z_{\vec{k}}=0$. They are $\vec{K}= \frac{4\pi}{3\sqrt{3}a} \hat{\vec{e}}_x$ and $\vec{K}' = -\vec{K}$ for the honeycomb lattice and $\vec{K}=0$ and $\vec{K}'=\frac{\pi}{a} \hat{\vec{e}}_x$ for the $\pi$-flux square lattice. Irrespective of the lattice, $|Z_{\vec{k}}|$ is always inversion-symmetric under $\vec{k} \to - \vec{k}$. However one may note that when the inversion symmetry is made about any of the Dirac points, e.g. $(\vec{K}+\vec{q}) \to (\vec{K}-\vec{q})$ this "local" property only keeps true for the $\pi$-flux square lattice.

Defining $\Phi_{\vec{k}}=-{\rm Arg}(Z_{\vec{k}})$, the eigenenergies are $\varepsilon_{\vec{k}s} = s\, \varepsilon_{\vec{k}} = s \sqrt{\delta^2+|Z_{\vec{k}}|^2}$ (where $s=\pm$ is the band index) and the corresponding eigenvectors are:
\begin{eqnarray}
\label{eigen+}
|\vec{k}+\rangle = \cos(\theta_{\vec{k}}/2) \, |\vec{k}\textsc{a}\rangle + e^{{\rm i} \Phi_{\vec{k}}}  \, \sin(\theta_{\vec{k}}/2) \,  |\vec{k}\textsc{b}\rangle,\\
|\vec{k}-\rangle = - e^{-{\rm i} \Phi_{\vec{k}}}  \, \sin(\theta_{\vec{k}}/2) \, |\vec{k}\textsc{a}\rangle +  \cos(\theta_{\vec{k}}/2) \,  |\vec{k}\textsc{b}\rangle, \label{eigen-}
\end{eqnarray}
with $\cos\theta_{\vec{k}} = \delta/\varepsilon_{\vec{k}}$. Identifying the Bloch waves $|\vec{k}\textsc{a}\rangle$ and $|\vec{k}\textsc{b}\rangle$ as up and down components $|\uparrow\rangle $ and $|\downarrow\rangle$, we define the pseudo-spin spinors:
\begin{eqnarray}
|u_{\vec{k}+}\rangle = \begin{pmatrix} \cos(\theta_{\vec{k}}/2) \\  \sin(\theta_{\vec{k}}/2) \, e^{{\rm i} \Phi_{\vec{k}}} \end{pmatrix}, \\
|u_{\vec{k}-}\rangle = \begin{pmatrix} -\sin(\theta_{\vec{k}}/2) \, e^{-{\rm i} \Phi_{\vec{k}}}\\  \cos(\theta_{\vec{k}}/2)  \end{pmatrix},
\end{eqnarray}
with $\langle u_{\vec{k}s}|u_{\vec{k}s'}\rangle = \delta_{ss'}$. One should note that the eigenenergies of the $\pi$-flux square lattice always display the inversion symmetry $\vec{k} \to -\vec{k}$, a feature which proves central when discussing the anti-CBS effect at long times.

Expanding $\mathcal{H}_{\vec{k}}$ around the Dirac points, i.e. $\vec{q}=\vec{k}$ for the $\pi$-flux square lattice and $\vec{q}=\vec{k}-\vec{K}$ for the honeycomb lattice ($qa\ll1$), we recover the Dirac Hamiltonian~[Eq.~(1) in main text] at lowest order in $qa$ with $c= 2Ja/\hbar$ and $c=3Ja/(2\hbar)$ for the square and honeycomb lattice respectively. The mass of the Dirac particles is defined as $\delta = mc^2$ and the corresponding spinor waves are obtained by setting $\Phi_{\vec{k}} = \varphi_{\vec{q}}+\pi$ where $\varphi_{\vec{q}}$ is the polar angle of $\vec{q}$. The dispersion relation become $\varepsilon_{\vec{q}} = \hbar c \sqrt{q^2+q^2_m}$ with $q_m=mc/\hbar$. When higher-order terms are considered, trigonal warping sets in for the honeycomb case and destroys the inversion symmetry $\vec{q} \to - \vec{q}$ (equivalently $\varphi_{\vec{q}} \to \varphi_{\vec{q}}+\pi$), implying $\varepsilon_{-\vec{q}} \neq \varepsilon_{\vec{q}}$, while this inversion symmetry keeps exact for the $\pi$-flux square lattice. This is clearly seen in the Taylor expansion of $|Z_{\vec{k}}|$ when the next order term is considered:
\begin{equation}
|Z_{\vec{k}}/J| \approx 2qa - \frac{\cos^4\varphi_{\vec{q}}+\sin^4\varphi_{\vec{q}}}{3} \, (qa)^3
\end{equation}
for the $\pi$-flux square lattice, while 
\begin{equation}
|Z_{\vec{k}}/J| \approx \frac{3qa}{2} + \frac{3(\cos\varphi_{\vec{q}}\sin^2\varphi_{\vec{q}}-\cos^3\varphi_{\vec{q}})}{8} \, (qa)^2
\end{equation}
for the honeycomb lattice.

\section{\label{sec:maximally_cross}Scattering vertex and maximally-crossed series}
The disorder potential reads $V = \sum_{i\in\textsc{a}} n_i + \sum_{i\in\textsc{b}} n_i$. Since we assume Gaussian statistics, all its odd-order moments vanish and its even-order moments boil down, through Wick's theorem, to products of 2-point correlators. Thus the 2-point correlator is the only key ingredient in the diagrammatic expansion of the Dyson and Bethe-Salpeter equations.

A straightforward calculation show that:
\begin{align}
& \overline{\langle \vec{k}_2s_2|V|\vec{k}_1s_1\rangle\, \langle \vec{k}_3s_3|V|\vec{k}_4s_4\rangle} = \delta_{\vec{k}_1+\vec{k}_4,\vec{k}_2+\vec{k}_3} \nonumber \\ 
&\times P(\vec{k}_1-\vec{k}_2) \, \langle u_{\vec{k}_2s_2}|u_{\vec{k}_1s_1}\rangle \, \langle u_{\vec{k}_3s_3}|u_{\vec{k}_4s_4}\rangle,
\end{align}
where $P$ is the Fourier transform of the disorder 2-point spatial correlator. Since we assume here that the disorder correlation length is much larger than the lattice constant ($\zeta \gg a$), $P$ is peaked around $0$ and $\vec{k}_1 \approx \vec{k}_2$. In turn, because of momentum conservation, $\vec{k}_3 \approx \vec{k}_4$. This enforces $\langle u_{\vec{k}_2s_2}|u_{\vec{k}_1s_1}\rangle \langle u_{\vec{k}_3s_3}|u_{\vec{k}_4s_4}\rangle\approx \delta_{s_1s_2}\delta_{s_3s_4}$, implying that the band index is an approximate good quantum number. Similar considerations show that the self-energy is also quasi-diagonal in the band index, at least at the Born approximation.

\begin{figure}
  \includegraphics[width=0.4\textwidth]{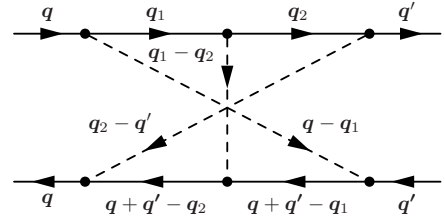}\caption{\label{fig:3mc} 3rd-order maximally-crossed diagrams for incoming and outgoing wave vectors $\vec{q}$ and $\vec{q}'$. Full circles connected by a dashed line correspond to a scattering vertex. Solid lines between two consecutive full circles represent average Green's function $\overline{G}$ solving Dyson equation (retarded for the upper lines, advanced for the lower lines). At each vertex, momentum conservation is applied. As one can easily see if $\vec{q}_i$ ($i=1,2$ here) is an intermediate momentum for the upper line, the corresponding momentum in the lower line is $\vec{q}'_i = \vec{q}+\vec{q}'-\vec{q}_i$. This means that reciprocal paths connecting $\vec{q}$ to $\vec{q}'$ are thus images of each other through an inversion about $(\vec{q}+\vec{q}')/2$.
  }
\end{figure}

As an illustrative example, Figure \ref{fig:3mc} shows the 3rd-order term in the maximally-crossed series for incoming and outgoing Dirac momenta $\vec{q}$ and $\vec{q}'$. The band index being approximately conserved here, it is not specified in the figure. Applying momentum conservation at each scattering vertex, it is easy to show that an intermediate Dirac momentum $\vec{q}_i$ in one scattering path is mapped onto $\vec{q}'_i = \vec{q}+\vec{q}'-\vec{q}_i$ in the reciprocal path. Reciprocal scattering paths connecting $\vec{q}$ to $\vec{q}'$ in momentum space are images of each other through the inversion about $(\vec{q}+\vec{q}')/2$, see Fig.~2 in main text. 

\section{\label{sec:dynamicFac}Dynamical factor}
With the change of variables $t_i \to y_i=t_i/t$, the dynamical factor [Eq.~(4) in main text] writes
\begin{align}
D_N(t) = t^{N}\int \!\!\cdot\cdot\cdot \!\! \int_0^{1}\big(\prod_{i=0}^{i=N}dy_i\big) \, \delta\big(\sqrt{N+1}\vec{v}\cdot
\vec{y}-1\big) \,  \Exp{-\mri \vec{x}\cdot\vec{y}},
\end{align}
where $\vec{x} = (\omega_0t, \cdot\cdot\cdot, \omega_Nt)$ and $\vec{v}=\frac{1}{\sqrt{N+1}}(1,\ldots,1)$ is the hyper-cube diagonal unit vector. We write $\vec{y}=y_\parallel \vec{v}+\vec{r}$ and $\vec{x}=x_\parallel \vec{v} + \vec{x}_\perp$, where $\vec{r}$ and  $\vec{x}_\perp$ are orthogonal to $\vec{v}$. We have $x_\parallel = \sqrt{N+1}E_Nt/\hbar$ where $E_N = \frac{1}{N+1} \, \sum_{j=0}^N \varepsilon_{\vec{q}_i}$. Integrating out the delta distribution, one easily gets
\begin{equation}
D_N(t) = \frac{t^NV_N\Exp{-\mri E_Nt/\hbar}}{\sqrt{N+1}} \, \int_{\Omega_N} \frac{\big(d\vec{r}\big)}{V_N} \, \Exp{-\mri \vec{x}_\perp
\cdot\vec{r}}.
\end{equation}
The integral runs over the regular $N$-simplex $\Omega_N$ obtained as the intersection of the $(N+1)$-hypercube with the hyperplane orthogonal to $\vec{v}$ located at $y_\parallel = \frac{1}{\sqrt{N+1}}$, $(d\vec{r})$ is the N-volume element while $V_N = \frac{\sqrt{N+1}}{N!}$ is the volume of the $N$-simplex.

\begin{figure}[!ht]
  \includegraphics[width=0.5\textwidth]{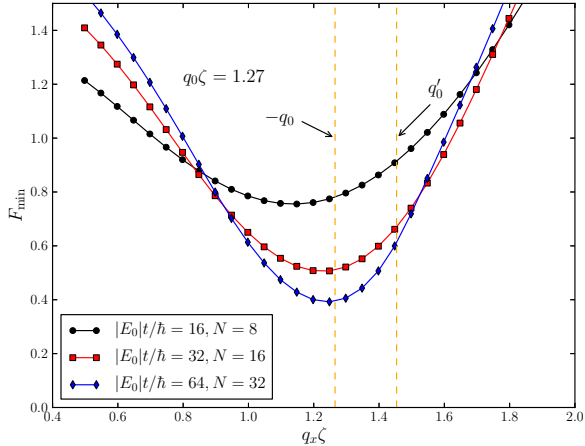}\caption{\label{fig:ACX_E0.4g0_F} Plot of the minimized function $F_{\rm min}$ 
as a function of the final momentum along the $x$-axis (in units of $1/\zeta$) for the honeycomb lattice and at $q_0\zeta=1.27$. The orange vertical dashed lines mark the momenta $\vec{q}=-\vec{q}_0$ and $\vec{q}=\vec{q}'_0$ (on-shell momentum opposite to $\vec{q}_0$). Noticeably, the minimum of $F_{\rm min}$ is located at $\vec{q}=-\vec{q}_0$, not at $\vec{q}'_0$, and is also the momentum where a density dip develops for $n_D(\vec{q},t)$, see left panels of Fig.~4 in the main text.
  }
\end{figure}

For $\vec{x}_\perp^2\ll(N+1)(N+2)$, we numerically find that
\begin{align}
\int_{\Omega_N} \frac{\big(d\vec{r}\big)}{V_N} \, \Exp{-\mri \vec{x}_\perp
\cdot\vec{r}} \approx\Exp{-\frac{\vec{x}_\perp^2}{2(N+1)(N+2)}}.
\end{align} 
Since $\vec{x}_\perp^2 =  \vec{x}^2-x_\parallel^2= (N+1) \sigma_E^2t^2/\hbar^2$, we thus get:
\begin{equation}
D_N(t) \approx \frac{t^N}{N!}\Exp{-\mri E_Nt/\hbar} \Exp{-\frac{\sigma_E^2t^2}{2\hbar^2(N+2)}},
\end{equation}
where the energy spread reads
\begin{equation}
\label{eq:energy_spread}
\sigma_E^2=\frac{1}{N+1}\sum_{i=0}^{i=N} \varepsilon_{\vec{q}_i}^2-E_N^2.
\end{equation}
We conclude that, for a given scattering path, (i) the dynamical phase is approximated by $(-E_N t/\hbar)$, and (ii) the dynamical factor decays at a rate determined by the spread of the intermediate energies.

\section{\label{sec:optimize}Optimization procedure}

\begin{figure}[!ht]
  \includegraphics[width=0.5\textwidth]{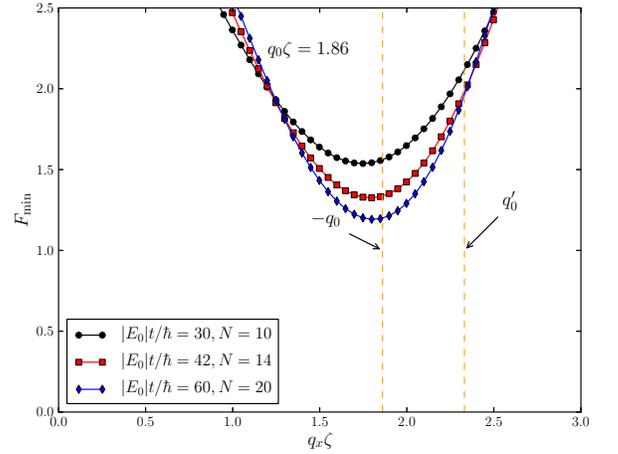}\caption{\label{fig:ACX_E0.6g0_F} Same as Fig.~\ref{fig:ACX_E0.4g0_F}, but for $q_0\zeta=1.86$. In this case, the energy shell is further away from the Dirac point and the trigonal warping is a more prominent effect. Here again the minimum is located at $-\vec{q}_0$ which is where the density dip is witnessed, see the right panels of Fig.~4 in the main text.
  }
\end{figure}
Within the previous approximation, we get
\begin{align}
&n_C(\vec{q},t) \approx e^{-\frac{t}{\tau_s}} \, \sum_N \frac{t^{2N}}{(N!)^2} \,I_N(\vec{q},t),\\
&I_N(\vec{q},t)= \int [d\mu] \,  e^{-\frac{({\sigma'_E}^2+\sigma^2_E)t^2}{2(N+2)\hbar^2}}\,  \cos(\Delta_N t -\Phi^{(N)}_g).
\end{align}
We infer the interference peak location from the condition $\cos(\Delta_N t -\Phi^{(N)}_g)=1$ obtained with the path $\{\vec{q}_0, \vec{q}_1, \cdot\cdot\cdot, \vec{q}_{N-1}, \vec{q}_N=\vec{q}\}$ maximizing the weight of the interference term with respect to the intermediate momenta $\vec{q}_i\quad(i=1,\ldots,N-1)$. This is achieved by maximizing the product $F$ of disorder correlation functions in the integration measure $[d\mu]$ with the Gaussian factor related to the energy spreads:
\begin{equation}
F = \frac{\zeta^2}{2}\sum_{i=0}^{i=N-1}(\vec{q}_{i+1}-\vec{q}_i)^2 + \frac{(\sigma_{E}^2+{\sigma'_E}^2)t^2}{2(N+2)\hbar^2}. \label{eq:F}
\end{equation}
Here $\sigma_E$ and ${\sigma'_E}$ are the energy spreads along the direct path and its reciprocal partner, see Eq.~\eqref{eq:energy_spread}. As we only expect a crude estimate for the peak position, we have further assumed $\vec{q}$ to lie on the energy shell. Results presented in Fig.~3 of the main text have been obtained using $N\approx t/\tau_s$, where the scattering mean free time $\tau_s$ is estimated from the time decay of the ballistic component. For smaller times, the path would be pulled slightly towards the Dirac point near $\vec{q}_0$ and $\vec{q}$ because the disorder correlation function then plays a more significant role.

\section{\label{sec:CBSpoint}CBS point and trigonal warping}
For each $\vec{q}$ lying on the $x$-axis, we minimize the function $F$, defined in Eq.~\eqref{eq:F}, with respect to the intermediate momenta $\vec{q}_i$ as explained in the previous section. For the parameters investigated in this work, the minimized function $F_{\rm min}$ always shows a minimum at $\vec{q}=-\vec{q}_0$, see Figs.~\ref{fig:ACX_E0.4g0_F} and~\ref{fig:ACX_E0.6g0_F}. Since the density dip in $n_D$ also occurs at $\vec{q}=-\vec{q}_0$, we conclude that, even in the presence of trigonal warping, $-\vec{q}_0$ remains the CBS momentum. We also note that $F_{\rm min}$ assumes larger values when $q_0\zeta$ gets larger.
%This observation is consistent with Fig.~\ref{fig:ring}, which shows that the CBS dip vanishes at a faster rate for $q_0\zeta=1.86$ than for $q_0\zeta=1.27$.
This observation is consistent with Fig.~4 in the main text, which shows that the CBS dip vanishes at a faster rate for $q_0\zeta=1.86$ than for $q_0\zeta=1.27$.
\bibliography{prlRing}
\end{document}